\documentclass[preprint,showpacs,preprintnumbers,amsmath,amssymb]{revtex4}
\usepackage{graphicx}
\begin{document}

\title{Slave particle approach to the finite temperature properties of
 ultracold Bose gases in optical lattices}

\author{Xiancong Lu$^1$, Jinbin Li$^2$ and Yue Yu$^1$ }
\affiliation{1. Institute of Theoretical Physics, Chinese Academy
of Sciences, P.O. Box 2735, Beijing 100080, China\\
2. Institute of
Physics, Chinese Academy of Sciences, Beijing 100080, China}
\date{\today}

\begin{abstract}
By using slave particle (slave boson and slave fermion) technique
on the Bose-Hubbard model, we study the finite temperature
properties of ultracold Bose gases in optical lattices. The phase
diagrams at finite temperature are depicted by including different
types of slave particles and the effect of the finite types of
slave particles is estimated. The superfluid density is evaluated
using the Landau second order phase transition theory. The atom
density, excitation spectrum and dispersion curve are also
computed at various temperatures, and how the Mott-insulator
evolves as the temperature increases is demonstrated. For most
quantities to be calculated, we find that there are no
qualitatively differences in using the slave boson or the slave
fermion approaches. However, when studying the stability of the
mean field state, we find that in contrast to the slave fermion
approach, the slave boson mean field state is not stable. Although
the slave boson mean field theory gives a qualitatively correct
phase boundary, it corresponds to a local maximum of Landau free
energy and can not describe the second order phase transition
because the coefficient $a_4$ of the fourth order term is always
negative in the free energy expansion.

\end{abstract}

\pacs{03.75.Lm,67.40.-w,39.25.+k}

\maketitle

\section{Introduction}

Ultracold atoms in optical lattices with highly tunable parameters
have provided an unique opportunity to simulate strongly
correlated phenomena in condensed matter physics
\cite{Bloch,gre,orzel}. The dynamics of such Bose atoms can be
described by a Bose-Hubbard model, and it was predicted that there
would be a quantum phase transition from the superfluid to the
Mott-insulator phase induced by varying the depth of the optical
potential \cite{Jaksch}. Recently, this phase transition has been
perfectly realized by Greiner et al. \cite{gre} by means of
loading $^{87}$Rb atoms into a three-dimensional optical lattices.
From then on the Bose-Hubbard model to the cold atomic gas in
optical lattices has received extensive studies.

The Bose-Hubbard model, with on-site interaction and tunnelling
between nearest neighbor sites, was used by Fisher et. al
\cite{fisher} to investigate the bosons in periodic and/or random
external potentials. There exist two types of phases, the
superfluid and Mott-insulating phases, in this model at an integer
lattice filling fraction and zero temperature. In the superfluid
phase, the tunnelling term dominates and all atoms occupy the
identical extended Bloch state. The ground state of this system
can be well described by a macroscopic wave function with
long-range phase coherence. In the Mott-insulating phase, the
interaction dominates and the ground state of the system instead
consists of the localized atomic wave function without phase
coherence \cite{gre}. This Mott-insulating phase can be
characterized by an integer filling factor, the existence of a gap
for particle-hole excitation, and zero compressibility. By varying
the parameters such as the density and the external potential, the
system would undergo a quantum phase transition and evolve from
the superfluid phase to Mott-insulating phase. In the past,
various theoretical approaches have been used to investigate this
superfluid/Mott-insulator transition at zero temperature such as
the strong-coupling expansion
\cite{freericks,freericks2,seng,Elstner}, Gutzwiller projection
ansatz \cite{Jaksch,Rokhsar,Krauth,Schroll}, quantum Monte Carlo
simulations \cite{Wessel,numerical}, and other mean-field
approximations \cite{fisher,oosten,Sheshardi}. By comparison,
there are less studies focusing on the nonzero temperature
properties \cite{D,pv,yy,Sheshardi,finiteT,Konabe}, and this will
be the main topic of this paper.

Slave boson technique was developed to deal with the interacting
fermion systems for the convenient use of mean-field
approximations \cite{Kotliar}. It has been extended to study the
hard-core bosons on a lattice in a functional integral
representation by Ziegler \cite{ziegler}, in which there are only
two states per site: empty and singly occupations. A similar
formulation for interacting boson systems has been obtained by
Fr\'{e}sard \cite{fresard}. The analytical expression for the
Mott-insulating lobes was obtained and the density-density
correlation function was calculated. Recently, this approach has
been adopted by Dicherscheid et al. \cite{D} to investigate the
nonzero temperature behaviors of the ultracold atoms in optical
lattices. These authors showed that the calculated phase diagram
at zero temperature agrees well with the results by other mean
field methods \cite{freericks,oosten} and they further depicted
the phase diagrams at finite temperatures. The slave boson and
slave fermion approaches are equivalent in principle without any
approximation. However, the difference comes into view if some
mean field approximations are made in practical calculations.
Bearing this in mind, one of us with Chui \cite{yy} performed a
comparative study between the slave fermion and slave boson
approaches under the same mean-field approximation. It was found
that both approaches give the same qualitative phase diagram, but
the quantitative behaviors by slave fermion approach are more
accurate. This may be related to the Fermi statistics which
automatically excludes two same types of slave fermions from
occupying the same site when the constraint, one site can be
occupied by only one slave particle, is relaxed in the mean field
approximation.

In this paper, we will extend the slave particle technique to
investigate the various finite temperature properties of the
Bose-Hubbard model. We will study the finite temperature phase
diagram, atom density, superfluid density, and excitation spectrum
for various temperatures and system parameters. The comparison
between slave fermion and slave boson approaches will be
performed. We show that although the mean field theory of slave
boson gives a qualitatively reasonable phase boundary, the
negative coefficient $a_4$ of fourth order term in the Landau free
energy expansion means that the zero of the derivative of the free
energy is a local maximum and thus the slave boson mean field
state is not stable, i.e., the slave boson mean field theory is
unable to describe the second order phase transition of the normal
liquid/superfluid. This shortcoming of the slave boson approach
does not exist in the slave fermion approach. In this sense, using
the slave fermion approach to study the strongly correlated Bose
atoms may be preponderant.

This paper was organized as follows. In Sec. \ref{2} we review the
basic formalism of the slave particle technique to the
Bose-Hubbard model. In Sec. \ref{3} we focus on the critical
temperature of superfluid-normal phase transition, and depict the
phase diagram. The atom density and the compressibility are
calculated in Sec. \ref{4}. In Sec. \ref{5}, the superfluid
density is determined and the stability of the mean field states
of the slave particles are discussed. Sec. \ref{6} is devoted to
calculate the excitation spectrum. Conclusions are given in Sec.
\ref{7}. In Appendix A, the details of the calculation of the
coefficient $a_4$ is provided.

\section{Slave particle formalism of Bose-Hubbard model}\label{2}

\subsection{Slave particle technique for a bosonic system}

For a bosonic system, the creation operator $a_{i}^{\dag}$ and
annihilation operator $a_{i}$ on site $i$ can be defined as
follows in the occupation-number representation:
\begin{eqnarray}
&&a_{i}^{\dag}|\alpha\rangle_{i}=\sqrt{\alpha+1}|\alpha+1\rangle_{i},
\nonumber\\
&&a_{i}|\alpha\rangle_{i}=\sqrt{\alpha}|\alpha-1\rangle_{i}.
\end{eqnarray}
They obey the basic boson commutation relation:
$[a_{i},a_{j}^{\dag}]=\delta_{ij}$. The state $|\alpha\rangle_{i}$
is an eigenstate of the particle number operator
$N_{i}=a_{i}^{\dag}a_{i}$, which counts the number of bosons on
site $i$, with the eigenvalue $\alpha$. On a single lattice site,
the occupation number can be any non-negative integer. Thus the
boson creation and annihilation operators can be decomposed into
\begin{eqnarray}
&&a_i^\dag=|1\rangle_{ii}\langle0|+\sqrt{2}|2\rangle_{ii}\langle1|+\cdots=
\sum_{\alpha=0}\sqrt{\alpha+1}|\alpha+1\rangle_{ii}\langle\alpha|,\nonumber\\
&&a_i=|0\rangle_{ii}\langle1|+\sqrt{2}|1\rangle_{ii}\langle2|+\cdots=
\sum_{\alpha=0}\sqrt{\alpha+1}|\alpha\rangle_{ii}\langle\alpha+1|.
\end{eqnarray}
The justification of this decomposition is that it does satisfy
the original commutation relation
$[a_{i},a_{j}^{\dag}]=\delta_{ij}$. Next, we identify every
occupation state on a site as a type of slave particle, that is to
say, mapping $|\alpha\rangle_i$ and $_i\langle\alpha|$ to the
creation operator $a_{\alpha,i}^{\dag}$ and annihilation operator
$a_{\alpha,i}$ of the slave particle. In a slave fermion approach
these operators are forced to satisfy the fermion anticommutation
relation
$\{a_{\alpha,i},a_{\beta,j}^\dag\}=\delta_{\alpha\beta}\delta_{ij}$,
and in a slave boson approach they are forced to obey boson
commutation relation
$[a_{\alpha,i},a_{\beta,j}^\dag]=\delta_{\alpha\beta}\delta_{ij}$.
Then $a_{i}^{\dag}$ and $a_{i}$ can be rewritten as
\begin{equation}\label{transformation}
\begin{split}
a_{i}^{\dag}=\sum_{\alpha=0}\sqrt{\alpha+1}a_{\alpha+1,i}^{\dag}a_{\alpha,i},\\
a_i=\sum_{\alpha=0}\sqrt{\alpha+1}a_{\alpha,i}^{\dag}a_{\alpha+1,i}.
\end{split}
\end{equation}
When inserting the above equations in $[a_{i},a_{j}^{\dag}]$, one
finds that, only when the constraint
\begin{equation}\label{constraint}
\sum_{\alpha=0}n_i^{\alpha}=\sum_{\alpha=0}a_{\alpha,i}^{\dag}a_{\alpha,i}=1,
\end{equation}
is satisfied, could the original boson commutation relation
$[a_{i},a_{j}^{\dag}]=\delta_{ij}$ be reproduced by either slave
fermion or slave boson approaches. This implies that the slave
particle transformation (\ref{transformation}) along with the
constraint (\ref{constraint}) and
$\{a_{\alpha,i},a_{\beta,j}^\dag\}=\delta_{\alpha\beta}\delta_{ij}$
(slave fermion) or
$[a_{\alpha,i},a_{\beta,j}^\dag]=\delta_{\alpha\beta}\delta_{ij}$
(slave boson) can describe the whole physics of the original boson
system.

\subsection{Bose-Hubbard model and functional integral representation }

In the second quantization, the translationally invariant
many-body Hamiltonian of cold Bose gases confined by an external
optical lattice potential $V_{ext}(\textbf{r})$ is given
 by \cite{dalfovo} ,
\begin{equation}\label{field}
\begin{split}
H=&\int{d\textbf{r}\Psi^\dagger(\textbf{r})\left(-\frac{\hbar^2}{2m}\nabla^2
+V_{ext}(\textbf{r})
-\mu\right)\Psi(\textbf{r})}\\
&+\frac{1}{2}\int{d\textbf{r}d\textbf{r}'\Psi^\dagger(\textbf{r})
\Psi^\dagger(\textbf{r}')
V(\textbf{r}-\textbf{r}')\Psi(\textbf{r}')\Psi(\textbf{r})},
\end{split}
\end{equation}
where $\Psi(\textbf{r})$ ($\Psi^\dagger(\textbf{r})$) is the boson
field operator that annihilates (creates) a particle at the
position \textbf{r}, $V(\textbf{r}-\textbf{r}')$ is the two-body
interatomic potential, and $\mu$ is the chemical potential.  In
the case of a dilute cold atom gas, we can approximate
$V(\textbf{r}-\textbf{r}')$ with an effective interaction
$g\delta(\textbf{r}-\textbf{r}')$, where $g=4\pi{a_s}\hbar^2/m$
with $a_s$ the s-wave scattering length and $m$ the mass of the
atoms. When expanding the field operators in the Wannier basis and
keeping only the lowest vibrational states, namely
$\Psi(\textbf{r})=\sum_ia_iw(\textbf{r}-\textbf{r}_i)$,
 eq. (\ref{field}) can
be rewritten as the Bose-Hubbard Hamiltonian \cite{Jaksch}:
\begin{equation}\label{bhmodel}
H=-t\sum_{<ij>}a^\dag_ia_j-\mu\sum_in_i+\frac{U}{2}\sum_in_i(n_i-1),
\end{equation}
in which $n_i=a_i^\dag{a_i}$ is the particle number operator. The
symbol $\langle{ij}\rangle$ denotes the sum over all nearest
neighbor sites. $t$ and $U$ are the hopping amplitude and on-site
interaction, respectively,
\begin{eqnarray}
t&=&\int{d\textbf{r}w^*(\textbf{r}-\textbf{r}_i)\left(-\frac{\hbar^2}{2m}\nabla^2+
V_{ext}(\textbf{r})\right)w(\textbf{r}-\textbf{r}_j)},\nonumber\\
U&=&g\int{d\textbf{r}|w(\textbf{r})|^4}.
\end{eqnarray}
For small occupation per site, one can use the single particle
Wannier function to calculate these parameters. However, in the
case of the multi-occupation, these parameters have to be
calculated by considering the interaction broadening of the
Wannier function \cite{jinbin}. Substituting eq.
(\ref{transformation}) into eq. (\ref{bhmodel}), the Bose-Hubbard
Hamiltonian can be replaced by
\begin{equation}
\begin{split}
H=&-t\sum_{<ij>}\sum_{\alpha,\beta}\sqrt{\alpha+1}\sqrt{\beta+1}
a_{\alpha+1,i}^{\dag}a_{\alpha,i}a_{\beta,j}^{\dag}a_{\beta+1,j}\\
&-\mu\sum_i\sum_{\alpha}{\alpha}n_i^{\alpha}+\frac{U}{2}
\sum_i\sum_{\alpha}\alpha(\alpha-1)n_i^\alpha.
\end{split}
\end{equation}
Following the steps in Refs. \cite{negele,stoof}, we write the
partition function as an imaginary time coherent state path
integral:
\begin{eqnarray}
&&Z={\rm
Tr}e^{-{\beta}H}=\int{Da_{\alpha}D\bar{a}_{\alpha}D{\lambda}
e^{-S[\bar{a}_{\alpha},a_{\alpha},\lambda]}},\nonumber\\
&&S[\bar{a}_{\alpha},a_{\alpha},\lambda]=\int_0^{\beta}{d\tau}\biggl\{
\sum_i\sum_{\alpha}\bar{a}_{\alpha,i}
[\partial_{\tau}-\alpha\mu+\frac{U}{2}\alpha(\alpha-1)
-i\lambda_i]a_{\alpha,i}\nonumber\\
&&+i\sum_i\lambda_i-t\sum_{<ij>}\sum_{\alpha,\beta}\sqrt{\alpha+1}\sqrt{\beta+1}
\bar{a}_{\alpha+1,i}a_{\alpha,i}\bar{a}_{\beta,j}a_{\beta+1,j}\biggr\},
\end{eqnarray}
where $\bar{a}_{\alpha,i}$ and $a_{\alpha,i}$ are introduced as
ordinary complex numbers in the slave boson approach, and as
Grassmann variables in the slave fermion approach satisfying the
Grassmann algebra, i.e.,
$\{a_{\alpha},a_{\alpha}\}=\{a_{\alpha},\bar{a}_{\alpha}\}
=\{\bar{a}_{\alpha},\bar{a}_{\alpha}\}$, $\int da_{\alpha} 1=0$
and $\int da_{\alpha} a_{\alpha}=1$ \cite{negele,stoof}. The
gaussian integrals over them are
\begin{equation}\label{gaussian integral}
\int{D}a_{\alpha}D\bar{a}_{\alpha}\exp\left\{-\sum_{\alpha\beta}\bar{a}_{\alpha}
A_{\alpha\beta}a_{\beta}\right\}=(\det{A})^{\pm}=e^{{\pm}Tr[\ln{A}]},
\end{equation}
where $\pm$ correspond to integral over Grassmann variables or
ordinary complex numbers. The Lagrange multiplier field
$\lambda_i(\tau)$ comes from the constraint (\ref{constraint}),
namely, $\prod_i\delta(\sum_{\alpha}n_i^{\alpha}-1)$. The unit has
been set to $\hbar=k_B=1$ in all formulas. In order to decouple
the hopping term, a Hubbard-Stratonovich transformation is
performed by adding a complete square term to the action which
contributes to the partition function a constant,
\begin{equation}
\begin{split}
\int{D\Phi^*D\Phi}\exp&\left[-\int d\tau
t\sum_{<ij>}(\Phi_i^*-\sum_{\alpha}\sqrt{\alpha+1}
\bar{a}_{\alpha+1,i}a_{\alpha,i})\right.\\
&\left.\times(\Phi_j-\sum_{\alpha}\sqrt{\alpha+1}
\bar{a}_{\alpha,j}a_{\alpha+1,j})\right].
\end{split}
\end{equation}
Then the partition function is replaced by
\begin{equation}
\begin{split}
Z&=\int{D\Phi^*D\Phi D\bar{a}_{\alpha}Da_{\alpha}D{\lambda}
e^{-S_{eff}[\Phi,a_{\alpha},\lambda]}},\\
S_{eff}[\Phi,a_{\alpha},\lambda]&=\int_0^{\beta}{d\tau}\left\{
\sum_i\sum_{\alpha}\bar{a}_{\alpha,i}
\left[\partial_{\tau}-\alpha\mu+\frac{U}{2}\alpha(\alpha-1)
-i\lambda_i\right]a_{\alpha,i}\right.\\
&\left.+i\sum_i\lambda_i+t\sum_{<ij>}(\Phi_i^*\Phi_j-\Phi_i^*
\sum_{\alpha}\sqrt{\alpha+1}
\bar{a}_{\alpha,j}a_{\alpha+1,j}\right.\\
&\left.-\Phi_j\sum_{\alpha}\sqrt{\alpha+1}\bar{a}_{\alpha+1,i}a_{\alpha,i})\right\}.\\
\end{split}
\end{equation}
The Hubbard-Stratonovich field $\Phi_i$ introduced here can be
identified as the order parameter of superfluid for
$\langle\Phi_i\rangle=\langle\sum_{\alpha}\sqrt{\alpha+1}
\bar{a}_{\alpha,i}a_{\alpha+1,i})\rangle =\langle{a_i}\rangle$. We
then perform a Fourier transform on the field $A_i$:
\begin{equation}
\begin{split}
A_i=\frac{1}{\sqrt{L\beta}}\sum_{\textbf{k},n}A_{\textbf{k}n}e^{i(\textbf{k}
\cdot{\textbf{r}_i}-\omega_n\tau)},
\end{split}
\end{equation}
where $L$ is the total number of sites in the optical lattice,
$\omega_n$ is the Matsubara frequency, which equals
$(2n+1)\pi/\beta$ and $2n\pi/\beta$ for fermion and boson fields,
respectively. The effective action $S_{eff}$ now reads
\begin{equation}\label{action}
\begin{split}
S_{eff}[\Phi,a_{\alpha},\lambda]&=
\sum_{\textbf{k},n}\sum_{\alpha}\bar{a}_{\alpha,\textbf{k}n}\left[-i\omega_n-\alpha\mu
+\frac{U}{2}\alpha(\alpha-1)
\right]a_{\alpha,\textbf{k}n}+i\sqrt{L\beta}\lambda_{\textbf{0},0}\\
&-i\frac{1}{\sqrt{L\beta}}\sum_{\textbf{k},\textbf{q},n,n'}
\sum_{\alpha}\lambda_{\textbf{q},n'}
\bar{a}_{\alpha,\textbf{k}n}a_{\alpha,(\textbf{k+q})(n+n')}
+\sum_{\textbf{k},n}\epsilon_{\textbf{k}}|\Phi_{\textbf{k},n}|^2\\
&-\sum_{\textbf{k},\textbf{k}',n,n'}\sum_{\alpha}
\frac{\epsilon_{\textbf{k}'}}{\sqrt{L\beta}}\left\{\left(\sqrt{\alpha+1}
\bar{a}_{\alpha+1,(\textbf{k+k}')(n+n')}
a_{\alpha,\textbf{k}n}\right)\Phi_{{\textbf{k}',n'}}\right.\\
&\left.+{\left(\sqrt{\alpha+1}\bar{a}_{\alpha,\textbf{k}n}
a_{\alpha+1,(\textbf{k+k}')(n+n')}
\right)\Phi_{{\textbf{k}',n'}}^*}\right\},
\end{split}
\end{equation}
where $\epsilon_{\textbf{k}}=2t\sum_{i=1}^d\cos({k_i}a)$ with $d$
and $a$ being the dimension and spacing constant of the lattice.
So far, we have obtained an effective action in a functional
integral representation by the slave particle approach. It is a
reexpression of the original Bose-Hubbard model since all
transformations we made are rigorous. This effective action is the
starting point of our calculations.

\subsection{Perturbation theory}
The system can not be exactly solved with the constraint
(\ref{constraint}), which means there is exactly one type of slave
particle per site \cite{Kotliar,p.a.lee}. We then relax it to one
slave particle per site on average over the whole lattice. To
realize this, one can replace all $\lambda_{\textbf{k},n}$ in the
action (\ref{action}) with a constant $\lambda_{\textbf{0},0}$.
This approximation is widely used in dealing with the interacting
fermion systems \cite{Kotliar,p.a.lee}. It implies that the
multi-occupation of the slave particles on one site is allowed.
However, the behaviors of the slave boson and slave fermion are
slightly different in the relaxing process. The multi-occupation
of the same type of slave boson is allowed, while the
multi-occupation of the same type of slave fermion is forbidden by
the Pauli principle \cite{yy}. This will lead to some quantitative
differences in the results by these two approaches.

According to the Landau phase transition theory, the order
parameter $\Phi$ near the critical point is small and then the
perturbation theory can be used. In the following, we will try to
integrate the slave particle field out of the action
(\ref{action}), and perform perturbation calculation toward $\Phi$
along the way in Ref. \cite{D}. After relaxing the constraint
(\ref{constraint}), the action (\ref{action}) can be divided into
two parts:
\begin{equation*}
\begin{split}
S_{eff}[\Phi,a_{\alpha},\lambda]=S_0+S_I
\end{split}
\end{equation*}
\begin{equation*}
\begin{split}
S_0=iL\beta\lambda+\sum_{\textbf{k},n}\sum_{\alpha}
\bar{a}_{\alpha,\textbf{k}n}\left[-i\omega_n
+c(\alpha)\right]a_{\alpha,\textbf{k}n}
+\sum_{\textbf{k},n}\epsilon_{\textbf{k}}|\Phi_{\textbf{k},n}|^2=S_0^{sp}
+\sum_{\textbf{k},n}\epsilon_{\textbf{k}}|\Phi_{\textbf{k},n}|^2
\end{split}
\end{equation*}
\begin{equation*}
\begin{split}
S_I=-\sum_{\textbf{k},\textbf{k}',n,n'}&\sum_{\alpha}
\frac{\epsilon_{\textbf{k}'}}{\sqrt{L\beta}}\left\{\left(\sqrt{\alpha+1}
\bar{a}_{\alpha+1,(\textbf{k+k}')(n+n')}
a_{\alpha,\textbf{k}n}\right)\Phi_{{\textbf{k}',n'}}\right.\\
&\left.+{\left(\sqrt{\alpha+1}
\bar{a}_{\alpha,\textbf{k}n}a_{\alpha+1,(\textbf{k+k}')(n+n')}
\right)\Phi_{{\textbf{k}',n'}}^*}\right\},
\end{split}
\end{equation*}
in which $\lambda=\lambda_{\textbf{0},0}/\sqrt{L\beta}$ is a
constant and
\begin{eqnarray}\label{c(alpha)}
c(\alpha)=-i\lambda-\alpha\mu+\alpha(\alpha-1)\frac{U}{2}.
\end{eqnarray}
 The partition function $Z_0$ of
non-interacting slave particles comes from the contribution of the
zeroth-order term $S^{sp}_0$ and is given by
\begin{eqnarray}\label{z0}
Z_0=e^{-\beta\Omega_0}=\int{Da_{\alpha}D\bar{a}_{\alpha}e^{-S_0^{sp}}}
\end{eqnarray}
where $\Omega_0$ is the zeroth-order thermodynamic potential and
has the form:
\begin{eqnarray}\label{omega0}
-\Omega_0=i\lambda{L}\pm\frac{L}{\beta}\sum_{\alpha}\ln(1\pm
e^{-\beta c(\alpha)}),
\end{eqnarray}
where the $+(-)$ sign corresponds to the slave fermion (slave
boson), respectively. Then we can define the average of an
operator $A$ with respect to $S_0^{sp}$ as
\begin{equation}\label{average}
\begin{split}
\langle{A}\rangle_0=\frac{1}{Z_0}\int{Da_{\alpha}D\bar{a}_{\alpha}
A(a_{\alpha},\bar{a}_{\alpha})e^{-S_0^{sp}}}.
\end{split}
\end{equation}
For small $\Phi$, $e^{-S_{eff}}$ can be expanded in terms of $S_I$
\cite{D,stoof}, i.e.,
\begin{equation}
\begin{split}
e^{-S_{eff}}=e^{-(S_0+S_I)}\approx{e^{-S_{0}}\left[1-S_I+\frac{1}{2}S_I^2\right]}.
\end{split}
\end{equation}
After integrating out the slave particle field $a_{\alpha}$ and
$\bar{a}_{\alpha}$, we arrive at a new effective action
$S_{E,eff}$,
\begin{eqnarray}\label{e,eff}
e^{-S_{E,eff}}=\int{Da_{\alpha}D\bar{a}_{\alpha}e^{-S_{eff}}}
=e^{-\beta\Omega_0-\sum_{\textbf{k},n}\epsilon_{\textbf{k}}|\Phi_{\textbf{k},n}|^2}
\left[1-\langle{S_I}\rangle_0+\frac{1}{2}\langle{S_I^2}\rangle_0\right].
\end{eqnarray}
It is straightforward to calculate $\langle{S_I}\rangle_0$ and
$\langle{S_I^2}\rangle_0$ using the Wick's theorem and gaussian
integral formula (eq. (\ref{gaussian integral})),
\begin{eqnarray}
\langle{S_I}\rangle_0&=&0,\\
\langle{S_I^2}\rangle_0&=&\mp2\sum_{\textbf{k},\textbf{k}',n,n'}
\frac{\epsilon_{\textbf{k}'}^2
|\Phi_{{\textbf{k}',n'}}|^2}{L\beta}\sum_{\alpha}(\alpha+1)
\langle{\bar{a}_{\alpha,\textbf{k}n}a_{\alpha,\textbf{k}n}}\rangle_0\nonumber\\
&&\times\langle{\bar{a}_{\alpha+1,(\textbf{k+k}')(n+n')}
a_{\alpha+1,(\textbf{k+k}')(n+n')}}\rangle_0\nonumber\\
&=&\mp2\sum_{\textbf{k},\textbf{k}',n,n'}\frac{\epsilon_{\textbf{k}'}^2
|\Phi_{{\textbf{k}',n'}}|^2}{L\beta}\sum_{\alpha}(\alpha+1)
\frac{1}{-i\omega_n+c(\alpha)}\nonumber\\
&&\times\frac{1}{-i\omega_{n+n'}+c(\alpha+1)},
\end{eqnarray}
where $\mp$ correspond to the slave fermion and slave boson. After
performing the sums over Matsubara frequency $\omega_{n}$ and
$\textbf{k}$, we reduce $\langle{S_I^2}\rangle_0$ to
\begin{equation}
\begin{split}
\langle{S_I^2}\rangle_0=&2\sum_{\textbf{k},n}\epsilon_{\textbf{k}}^2
|\Phi_{{\textbf{k},n}}|^2\sum_{\alpha}(\alpha+1)\frac{n^{\alpha}-n^{\alpha+1}}
{-i\omega_n-\mu+\alpha{U}},
\end{split}
\end{equation}
where $n^{\alpha}$ is the occupation number and equal to
\begin{equation}\label{occupation number}
n^{\alpha}=\frac{1}{\exp\{\beta[-i\lambda-\alpha\mu+\alpha(\alpha-1)U/2]\}\pm1},
\end{equation}
in which the $+$ and $-$ sign correspond to slave fermion and
slave boson respectively. Because $\langle{S_I^2}\rangle_0$ is
small, we have
\begin{equation}
\begin{split}
\left[1-\langle{S_I}\rangle_0+\frac{1}{2}\langle{S_I^2}\rangle_0\right]
=\left[1+\frac{1}{2}\langle{S_I^2}\rangle_0\right]\approx
e^{\frac{1}{2}\langle{S_I^2}\rangle_0},
\end{split}
\end{equation}
and can rewrite eq. (\ref{e,eff}) as
\begin{eqnarray}
e^{-S_{E,eff}}
=e^{-\beta\Omega_0-\sum_{\textbf{k},n}\epsilon_{\textbf{k}}|\Phi_{\textbf{k},n}|^2
+\frac{1}{2}\langle{S_I^2}\rangle_0}.
\end{eqnarray}
Then the effective action of the order parameter field
$S_{E,eff}[\Phi^*,\Phi]$ may be obtained
\begin{equation}\label{e,eff2}
S_{E,eff}[\Phi^*,\Phi]=\beta\Omega_0
-\sum_{\textbf{k},n}\Phi^*_{\textbf{k},n}G^{-1}(\textbf{k},i\omega_n)\Phi_{\textbf{k},n},
\end{equation}
where the Green's function $G(\textbf{k},i\omega_n)$ is defined by
\begin{equation}\label{green}
-G^{-1}(\textbf{k},i\omega_n)=\epsilon_{\textbf{k}}-\epsilon_{\textbf{k}}^2\sum_{\alpha}
(\alpha+1)\frac{n^{\alpha}-n^{\alpha+1}}{-i\omega_n-\mu+\alpha{U}},
\end{equation}
which will be the starting point of our analysis.
\subsection{Mean-field approximation}
In terms of eq. (\ref{e,eff2}), the partition function can be
calculated by gaussian integral
\begin{equation}
\begin{split}
Z=&e^{-\beta\Omega}=\int{D\Phi D\Phi^*e^{-S_{E,eff}}}\\
=&\int{{D\Phi}D\Phi^*e^{-\beta\Omega_0+\sum_{\textbf{k},n}\Phi^*_{\textbf{k},n}
G^{-1}(\textbf{k},i\omega_n)\Phi_{\textbf{k},n}}}\\
=&e^{-\beta\Omega_0-\sum_{\textbf{k},n}\ln[-{\beta}G^{-1}(\textbf{k},i\omega_n)]},
\end{split}
\end{equation}
from which the thermodynamic potential $\Omega$ can be extracted:
\begin{eqnarray}\label{omega}
\Omega=\Omega_0+\frac{1}{\beta}\sum_{\textbf{k},n}\ln[-{\beta}G^{-1}(\textbf{k},i\omega_n)].
\end{eqnarray}
We then perform a saddle point approximation to the constraint
field $\lambda$, which means that we only choose the $\lambda$
minimizing the thermodynamic potential:
$\partial\Omega/\partial\lambda=0$. In addition, the particle
number conservation condition requires
$-\partial\Omega/\partial\mu=N$. When inserting eq. (\ref{omega})
into these two conditions, we have
\begin{equation}
\begin{split}
L\sum_{\alpha}(1-n^{\alpha})-\frac{i}{\beta}\sum_{\textbf{k},n}G(\textbf{k},i\omega_n)
\frac{{\partial}G^{-1}(\textbf{k},i\omega_n)}{\partial\lambda}=0,\\
L\sum_{\alpha}{\alpha}n^{\alpha}+\frac{1}{\beta}\sum_{\textbf{k},n}G(\textbf{k},i\omega_n)
\frac{{\partial}G^{-1}(\textbf{k},i\omega_n)}{\partial\mu}=N.
\end{split}
\end{equation}
The mean-field approximation means the last terms of the above
equations may be neglected. That is, all the fluctuations coming
from the Green's function would not be considered. Then the
following two equations can be derived,
\begin{equation}\label{constrain1}
\sum_{\alpha=0}n^{\alpha}=1,
\end{equation}
\begin{equation}\label{constrain2}
\sum_{\alpha=0}{\alpha}n^{\alpha}=\frac{N}{L}=n,
\end{equation}
where the $n=N/L$ is the average particle density. One can see
that eq. (\ref{constrain1}), which implies there is one slave
fermion per site on average, is exactly the relaxed constrain
(\ref{constraint}).

\section{The superfluid-normal phase transition}\label{3}

As we know, the quantum phase transition from superfluid to
Mott-insulator only occurs at $T=0$. At finite temperature, a
superfluid-normal phase transition will be induced
\cite{D,yy,pv,ziegler}. The Landau theory shows that, near the
critical point of superfluid-normal phase transition, the order
parameter of superfluid $\Phi$ is small and the Landau free energy
can be expanded in terms of it. The critical point can be
determined by the coefficient of the second order term $|\Phi|^2$,
which is $G^{-1}(\textbf{0},0)=0$ in our case \cite{D,yy}.
According to eq. (\ref{green}) and noting that $\epsilon_0=zt$, we
have
\begin{equation}\label{transitionpoint}
\sum_{\alpha=0}(\alpha+1)\frac{n^{\alpha+1}-n^{\alpha}}{\bar{\mu}-{\alpha}\bar{U}}=1,
\end{equation}
where $\bar{\mu}=\mu/zt$ and $\bar{U}=U/zt$ are dimensionless
chemical potential and on-site repulsion strength with $z$ being
the number of nearest neighbors. The dimensionless critical
temperature $\bar{T}_c=T_c/zt$ can be obtained by solving eqs.
(\ref{occupation number}), (\ref{constrain1}), (\ref{constrain2}),
and (\ref{transitionpoint}) together. Note that our results have a
clear mean field nature and the dimensionality of the lattice
appears only as a numerical factor. Because there are infinite
types of slave particles, a cut-off toward $\alpha$ should be made
when solving these equations. It has been shown by one of us and
Chui \cite{yy} that, for a large $\bar{U}$ a small cut-off (e.g.
$\alpha_M=3$) is a good approximation, but in the region of small
$\bar{U}$ the contributions from large $\alpha$ slave particles
cannot be neglected and a larger $\alpha_M$ is required to obtain
a quantitatively reliable result. In this work, we will work at a
relative large $\alpha_M$ ( say, 9) to obtain a more accurate
phase diagram of superfluid-normal transition and estimate the
effect of the finite types of slave particles.
\begin{figure}
\begin{center}
\includegraphics[width=0.6\columnwidth]{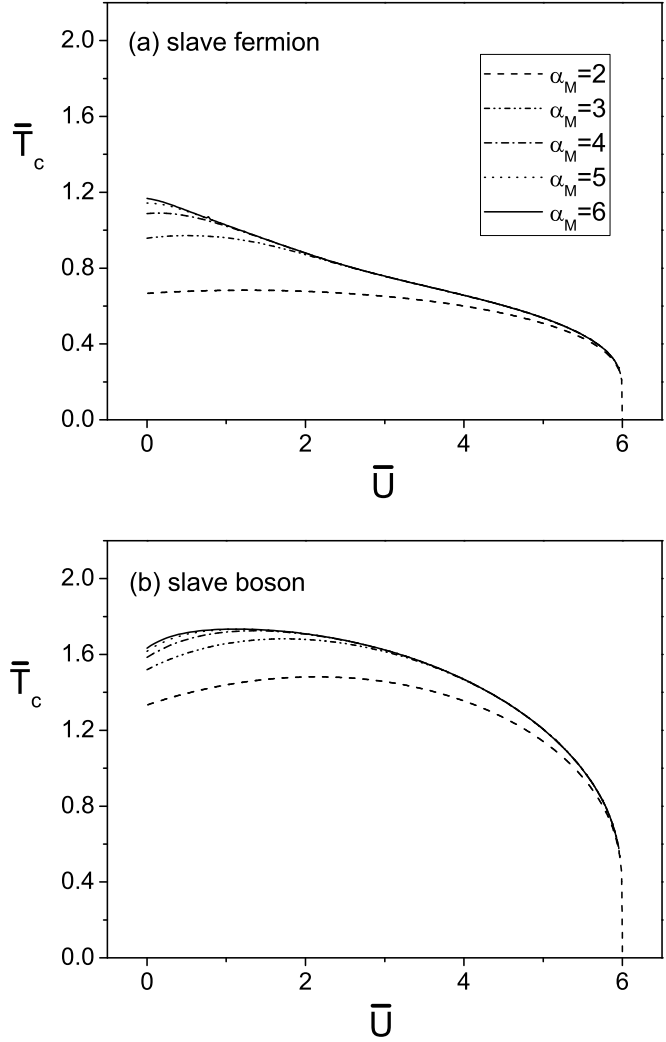}\\
\caption{\label{Tc-U}The superfluid-normal phase diagrams with
various cut-off $\alpha_M$ in the slave fermion ((a)) and the
slave boson ((b)) approaches. The average density is set to be
$n=1$. $\bar{T}_c=T_c/zt$ and $\bar{U}=U/zt$ are the dimensionless
critical temperature and on-site interaction, respectively.}
\end{center}
\end{figure}

\begin{figure}
\begin{center}
\includegraphics[width=0.6\columnwidth]{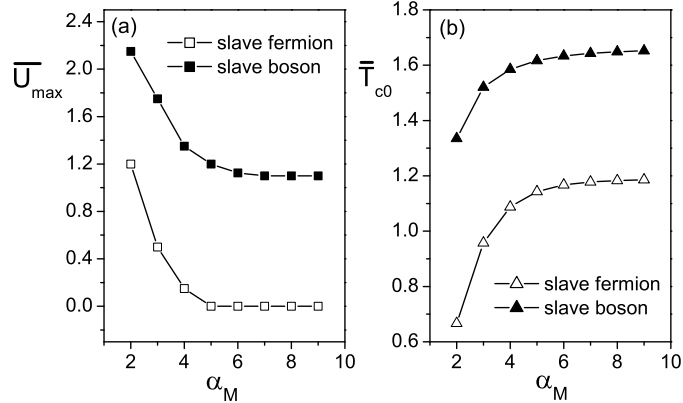}\\
\caption{\label{am} The position of local maximum $\bar{U}_{max}$
(panel (a)) and the critical temperature $\bar{T}_{c0}$ at
$\bar{U}=0$ (panel (b)) as a function of the cut-off $\alpha_M$
with the average density $n=1$.}
\end{center}
\end{figure}

When considering the commensurate state and restricting the
average density to $n=N/L=1$, we obtain the superfluid-normal
phase diagrams with various cut-off $\alpha_M$ in both slave
fermion and slave boson approaches. It can be seen from Fig.
\ref{Tc-U} that, in both approaches, the curves of various cut-off
are different from each other in the small $\bar{U}$ region. As
$\bar{U}$ increases, these differences become small and disappear
gradually when approaching the critical point of superfluid-Mott
insulator transition. This result gives a further support to the
statement of Ref. \cite{yy} that more types of slave particles
should be taken into account in the small $\bar{U}$ region.
Moreover, we find that the $\bar{T}_c-\bar{U}$ curves with
$\alpha_M>7$, which are not shown in Fig. \ref{Tc-U}, almost
coincide with the curve with $\alpha_M=6$. This implies that even
in the small $\bar{U}$ region the contributions from the slave
particles with $\alpha>6$ can be reasonably neglected. To make the
effect of finite cut-off $\alpha_M$ clearer, we plot the position
of local maximum in the $\bar{T}_c-\bar{U}$ curve, denoted by
$\bar{U}_{max}$, and the critical temperature at $\bar{U}=0$,
denoted by $\bar{T}_{c0}$, as a function of $\alpha_M$ in Fig.
\ref{am}. We can see that, as the cut-off $\alpha_M$ increases,
the positions of local maximum $\bar{U}_{max}$ move to small
$\bar{U}$ and approach steadily to 0 and 1.1 after $\alpha_M>6$
for the slave fermion and the slave boson, respectively. The
critical temperatures $\bar{T}_{c0}$ shown in Fig. \ref{am}(b)
initially increase largely when we increase the $\alpha_M$, but
their dependencies on it become very small after $\alpha_M>6$ in
both slave particle approaches. All of these behaviors show that,
in the case of low density (e.g. $n=1$ here), the finite cut-off
such as $\alpha_M=6$ is a good approximation in all range of
$\bar{U}$.

We now make a comparison between the phase diagrams in Fig.
\ref{Tc-U} by these two slave particle approaches. One can see
that the $\bar{T}_c$ of the slave fermion is always smaller than
that of the slave boson. As mentioned in Refs. \cite{D} and
\cite{yy}, the local maximum in the $\bar{T}_c-\bar{U}$ curve is
unreasonable. We can see from Fig. \ref{am}(a) that, when making a
cut-off such as $\alpha_M=5$, the position of the local maximum in
the slave fermion picture moves to zero, which means the local
maximum disappears. However, it seems that the local maximum in
the slave boson picture could not be eliminated by merely making a
large cut-off. There is still a maximum located around
$\bar{U}_{max}=1.1$ even when a large cut-off such as $\alpha_M=9$
is made. Moreover, the position of the maximum almost does not
depend on the cut-off when $\alpha_M>6$. This unsatisfactory
feature may come from the approximation of relaxing the constrain
of eq. (\ref{constraint}), which has less severe effect on the
slave fermion than on the slave boson \cite{yy}. On the other
hand, the critical temperature $\bar{T}_{c0}$ at $\bar{U}=0$
should be identical to the Bose-Einstein condensation temperature
$\bar{T}_c^{ideal}$ of ideal Bose gas, which in the case of three
dimensions can be determined by \cite{dalfovo,Kleinert}
\begin{eqnarray}\label{Tc of ideal bose gas}
\bar{T}_c^{ideal}=\frac{2\pi}{3}\left[\frac{n}{g_{3/2}(0)}\right]^{2/3}
\end{eqnarray}
with $n$ being the average particle density and
$g_i(x)=\sum_{m=1}^{\infty}e^{mx}/m^i$. We have
$\bar{T}_c^{ideal}=1.10$ when $n=1$ \cite{note1}. As shown in Fig.
\ref{am}(b), $\bar{T}_{c0}$ approaches 1.18 and 1.66 for the slave
fermion and the slave boson, respectively. Therefore, in this
special case of $n=1$, $\bar{T}_{c0}$ of the slave fermion is more
appropriate for it is closer to the expected value 1.10. As shown
in the phase diagrams, both slave particle approaches yield the
critical interaction $\bar{U}_c=6$ when $\alpha_M=2$, which
slightly deviates from the well-known mean field value
$\bar{U}_c=5.83$. This can be attributed to the mean field
approximation of eqs. (32) and (33) and the finite cut-off of
$\alpha_M$ \cite{yy,D}. However, a remedy to the mean field
conditions is very difficult \cite{yy,D} and, when including more
types of slave particles, the calculation of $\bar{T}_c$ at very
low temperature is very hard too for the divergence and the
multi-solutions at $\bar{U}>6$ \cite{yy}. Hence, when
$\alpha_M>2$, we actually do not work out $\bar{T}_c$ in the very
low temperature region in this work \cite{note2}.

\begin{figure}
\begin{center}
\includegraphics[width=0.6\columnwidth]{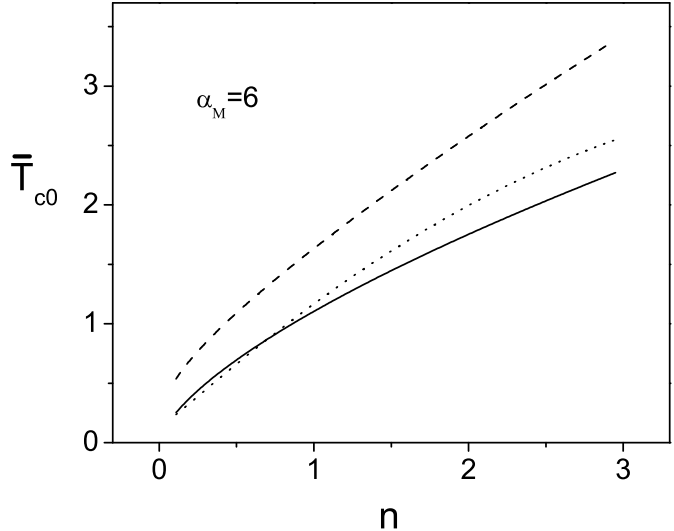}\\
\caption{\label{tc0-n} The critical temperature $\bar{T}_{c0}$ at
$\bar{U}=0$ as a function of average density $n$ with the cut-off
$\alpha_M=6$. The dashed and dotted lines correspond to the slave
boson and slave fermion, respectively. The solid line is the
critical temperature of ideal Bose gas in three dimensions.}
\end{center}
\end{figure}

All the critical temperatures calculated above are concentrated on
the commensurate state with fixed integer density (i.e. $n=1$),
which is important only to the exact quantum phase transition at
$T=0$. We next turn to investigate the $\bar{T}_{c0}$ away from
the integer filling and compare them with the critical temperature
of ideal Bose gas. In Fig. \ref{tc0-n}, the dependence of
$\bar{T}_{c0}$ on average density $n$ is plotted by both slave
particle approaches, and the critical temperature of
three-dimensional ideal Bose gas is plotted by eq. (\ref{Tc of
ideal bose gas}). We can see that the slave fermion curve is
always below the slave boson curve and closer to that of ideal
Bose gas. On the other hand, the deviations between both slave
particle curves and the ideal gas curve become larger as the
density increases. As mentioned above, the multi-occupation of
slave particles on one site is allowed in mean field
approximation. In the high density region, the multi-occupation
may occur more frequently for one should take more types of slave
particles into account. This is why the deviations from the ideal
gas become large in this region. However, in the slave fermion
approach, multi-occupation of the same type of slave fermion is
excluded, which makes this approximation less severe than in the
slave boson approach and gives a curve closer to that of the ideal
gas. One may notice that there is an unsatisfied feature in the
slave fermion picture too. The slave fermion curve and the ideal
Bose gas curve cross at two different values of the filling, which
implies that the slave fermion can not give the correct function
dependence of $\bar{T}_{c0}$ on $n$. In the end, we would pay some
attention to the mean field nature of our theory. As we know,
$\bar{T}_{c0}$ is calculated on base of the mean field
approximation and hence applicable to any spatial dimensions. This
will lead to the conclusion that the Bose-Einstein condensation
occurs in one and two-dimensional ideal Bose gas at finite
temperature, which is obviously wrong for the violation of the
Hohenberg theorem \cite{pitaevskii}. However, this is a well known
flaw: mean field theory usually breaks down in low dimensions due
to the large quantum fluctuation \cite{pitaevskii}.

\begin{figure}
\begin{center}
\includegraphics[width=0.6\columnwidth]{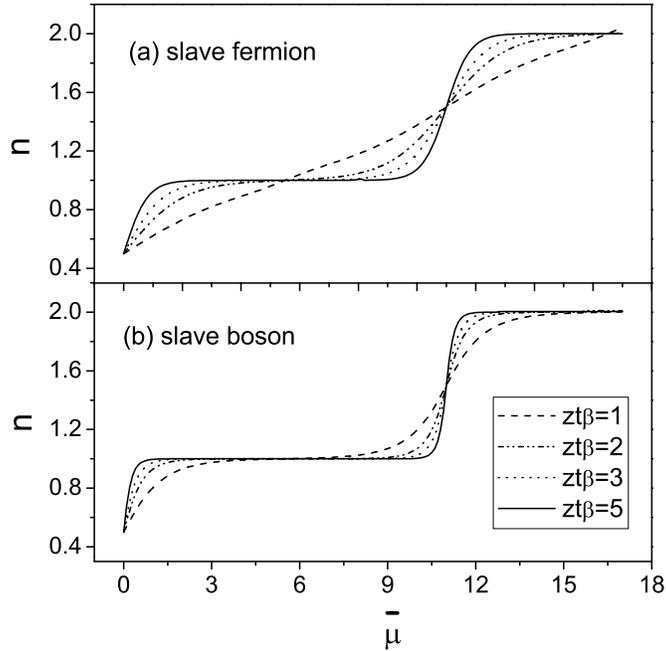}\\
\caption{\label{density} The average particle density $n$ as a
function of $\bar{\mu}$ with $\bar{U}=11$ at various temperatures,
obtained by the slave fermion (Fig. \ref{density}(a)) and slave
boson (Fig. \ref{density}(b)) approaches. The $\bar{\mu}$ and
$\bar{U}$ are the dimensionless chemical potential and on-site
interaction, respectively. Six types of slave particles (from
$n^0$ to $n^5$) are taken into account in the calculation. The
compressibility can be read out from the slope of the curve.}
\end{center}
\end{figure}

\section{Density and compressibility}\label{4}
The central features of the Mott-insulator at $T=0$ are the
integer filling factor $n^{\alpha}=\delta_{\alpha,\alpha'}$ and
the zero compressibility $\kappa=\partial{n}/\partial{\mu}=0$. At
finite temperature, there is no exact Mott-insulating phase
because the local filling factor may deviate from an integer and
the compressibility $\kappa$ does not equal zero. We will
investigate the finite temperature properties of these parameters
in this section. In principle, when given the value of $\bar{U}$,
one can calculate every type of occupation numbers $n^{\alpha}$ as
a function of $\bar{\mu}$ by combining eq. (\ref{occupation
number}) with eq. (\ref{constrain1}). Then the average particle
density $n$, which is a function of $\bar{\mu}$ too, can be
obtained according to eq. (\ref{constrain2}). After depicting the
$n-\bar{\mu}$ curve, one can read out the compressibility from the
slope of the curve. In Fig. \ref{density}, we show the calculated
$n-\bar{\mu}$ curves with $\bar{U}=11$ at different temperature in
both slave fermion and slave boson pictures. As the figure shows,
there are "steps" on the curves at very low temperature, where the
density is very close to an integer and the compressibility is
almost equal to zero. It is therefore reasonable to call these
regions the "Mott-insulator" at finite temperature \cite{D}. We
can see that the Mott-insulating regions in both slave particle
pictures diminish and disappear gradually as the temperature is
increased. However, the quantitative behaviors of them are a
little different, i.e., at the same temperature, the
compressibility of the slave fermion deviates from zero more
greatly than that of the slave boson. This implies that the
influence of temperature is greater on the slave fermion than on
the slave boson. In Fig. \ref{density}, there are only two "steps"
corresponding to $n=1$ and $n=2$. One can obtain the higher
"steps" by including more types of slave particles and working on
the larger $\bar{\mu}$. Note that, although the qualitative
behavior of this "step" structure is quite in accord with the zero
temperature density profile \cite{oosten}, the value of the
density deep in the superfluid region is unreliable for the
invalidity of perturbation theory.

\section{Superfluid density}\label{5}

\begin{figure}
\begin{center}
\includegraphics[width=0.6\columnwidth]{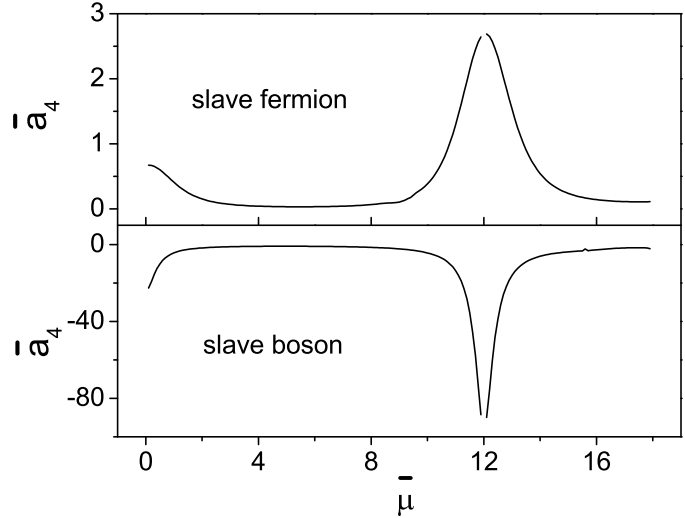}\\
\caption{\label{a4} The dimensionless fourth-order coefficient
$\bar{a}_4=L{\beta}a_4/zt$ as a function of chemical potential
$\bar{\mu}$ by the slave fermion and the slave boson approaches.
In the calculation, $\bar{U}=12$, $zt\beta=4$, and four types of
slave particles ($n^0$, $n^1$, $n^2$, and $n^3$) are taken into
account.}
\end{center}
\end{figure}

\begin{figure}
\begin{center}
\includegraphics[width=0.6\columnwidth]{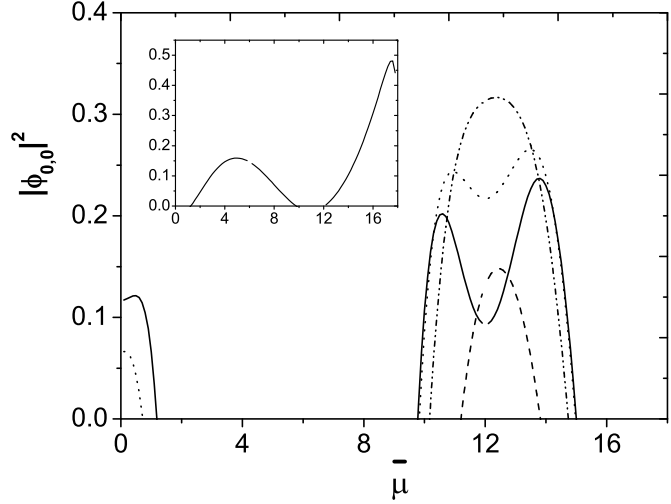}\\
\caption{\label{super} Superfluid density
$|\Phi_{\textbf{0},0}|^2$ as a function of chemical potential
$\bar{\mu}$ with $\bar{U}=12$ at different temperature. Four types
of slave particles ($n^0$, $n^1$, $n^2$, and $n^3$) are taken into
account. The solid, dotted, dashed-dotted-dotted, and dashed lines
are obtained by the slave fermion approach and correspond to
$zt\beta=$ 7, 4, 2.5, and 1.8, respectively. Inset shows the value
of $G(\textbf{0},0)/2a_4$ in the slave boson picture with
$zt\beta=7$, which should not be regarded as the superfluid
density because the $a_4$ is always negative.}
\end{center}
\end{figure}

As mentioned above, the action near the critical point can be
expended in powers of the order parameter $\Phi_{\textbf{0},0}$,
\begin{eqnarray}
S_{E,eff}=\beta\Omega_0-G^{-1}(\textbf{0},0)|\Phi_{\textbf{0},0}|^2
          +a_4|\Phi_{\textbf{0},0}|^4.
\end{eqnarray}
If the coefficient $a_4$ of fourth order term is positive, the
superfluid density $n_0$ can be determined by
\begin{eqnarray}
n_0=|\langle\Phi_{\textbf{0},0}\rangle|^2=\frac{G^{-1}(\textbf{0},0)}{2a_4},
\label{sd}
\end{eqnarray}
which minimizes the action. We calculate the coefficient $a_4$
following the steps in Ref. \cite{D} and show the detailed
calculation in Appendix \ref{appendix a}. The final result is
\begin{eqnarray}\label{a_4}
a_4=&-&\frac{{\epsilon_0}^4}{2L\beta}\sum_{\alpha}\left\{
\frac{(\alpha+1)^2\beta}{(\mu-{\alpha}U)^2}
[n^{\alpha}(1{\mp}n^{\alpha})+n^{\alpha+1}(1{\mp}n^{\alpha+1})]
+\frac{2(\alpha+1)^2}{({\alpha}U-\mu)^3}(n^{\alpha+1}-n^{\alpha})\right.\nonumber\\
&+&\left.\frac{2(\alpha+1)(\alpha+2)}{(\mu-{\alpha}U)^2[(2\alpha+1)U-2\mu]}n^{\alpha}
-\frac{2(\alpha+1)(\alpha+2)}{[\mu-(\alpha+1)U]^2[(2\alpha+1)U-2\mu]}
n^{\alpha+2}\right.\\
&-&\left.\frac{2(\alpha+1)(\alpha+2)U}{(\mu-{\alpha}U)^2[\mu-(\alpha+1)U]^2}n^{\alpha+1}
-\frac{2(\alpha+1)(\alpha+2)\beta}{(\mu-{\alpha}U)[\mu-(\alpha+1)U]}
n^{\alpha+1}(1\mp n^{\alpha+1})\right\},\nonumber
\end{eqnarray}
where $\mp$ correspond to the slave fermion or the slave boson
\cite{note} and $n^{\alpha}$ is defined in eq. (\ref{occupation
number}). In Fig. \ref{a4}, we plot the dimensionless
$\bar{a}_4=L{\beta}a_4/zt$ as a function of the chemical potential
$\bar{\mu}$ with $\bar{U}=12$ and $zt\beta=4$. We can see that the
sign difference in $a_4$ (i.e., $\mp$) leads to a very striking
result, i.e., $a_4$ of the slave boson is always negative. This
means the Landau free energy is not minimized but maximized at eq.
(\ref{sd}), that is, the mean field theory of the slave boson is
not stable and can not give the Landau second order phase
transition. The physical reason for this instability may come from
the condensation of the slave bosons due to the relaxation of the
constraint of one salve boson per site. To see this point more
explicitly, we depict in Fig. \ref{super} the dependence of the
superfluid density $|\Phi_{\textbf{0},0}|^2$ on the chemical
potential $\bar{\mu}$ at various temperature. The inset shows the
value of $G(\textbf{0},0)/2a_4$ as a function of $\bar{\mu}$ in
the slave boson picture. We can see that it is opposite to the
standard mean field theory at $T=0$ \cite{fisher,Sheshardi}, e.g.,
the "superfluid density" grows up at the places where are
Mott-insulating regions in the zero temperature phase diagram
\cite{note3}.

In the slave fermion approach, $a_4$ is positive definitely. Then
the Landau second order phase transition theory may be safely
applied. As shown in the Fig. \ref{super}, the superfluid regions
are consistent with those in the zero temperature phase diagram,
and become small and disappear gradually when the temperature is
increased. It is notable that the perturbation theory used here is
valid only near the critical point. Thus the value of the
superfluid density far away from the critical point may be
incorrect. The accurate value can be obtained by means of the
Bogoliubov theory \cite{dalfovo}, but this is beyond the scope of
this paper. Because only four types of slave fermions are
considered, the superfluid phase in the larger $\bar{\mu}$ region,
which corresponds to the higher filling factor, could not be
obtained in our calculation.

\iffalse Various physical properties can be calculated by the
slave particle technique. In terms of the discussions in this
section, we employ the slave fermion approach only in the rest of
the work.\fi

\section{The excitation spectrum}\label{6}

The excitation spectrum of the quasiparticle and quasihole can be
determined by the pole of the Green's function, that is, by the
equation $G^{-1}(\textbf{k},\omega)=0$. From eq. (\ref{green}), we
have
\begin{equation}\label{excitation}
\sum_{\alpha=0}(\alpha+1)\frac{n^{\alpha+1}-n^{\alpha}}{\omega+\mu-{\alpha}U}
=\frac{1}{\epsilon_{\textbf{k}}}.
\end{equation}
It is easy to show that both slave particle approaches give the
same excitation spectrum at $T=0$ for
$n^{\alpha}=\delta_{\alpha,\alpha'}$ with $\alpha'$ being an
integer filling factor. At nonzero temperature, one should take
more than one type of slave particle into account and eq.
(\ref{excitation}) will have more than two solutions \cite{D}. In
the case of low density $n=1$, it is reasonable to take only three
types of slave particles ($n^0$, $n^1$ and $n^2$) into account,
that is, only the processes in which the occupation of a site
changes among $n^0$, $n^1$ and $n^2$ are considered \cite{D}. By
this approximation, we can obtain two low-lying excitation spectra
analytically,
\begin{equation}\label{dispersion1}
\begin{split}
\omega_{\textbf{k}}^{\pm}+\mu&=\frac{U}{2}+\frac{1}{2}\epsilon_{\textbf{k}}
(2n^2-n^1-n^0)\\
&\pm\frac{1}{2}\sqrt{U^2+2(n^0-3n^1+2n^2)U\epsilon_{\textbf{k}}+(n^0+n^1-2n^2)^2
\epsilon_{\textbf{k}}^2} .
\end{split}
\end{equation}

\begin{figure}
\begin{center}
\includegraphics[width=0.6\columnwidth]{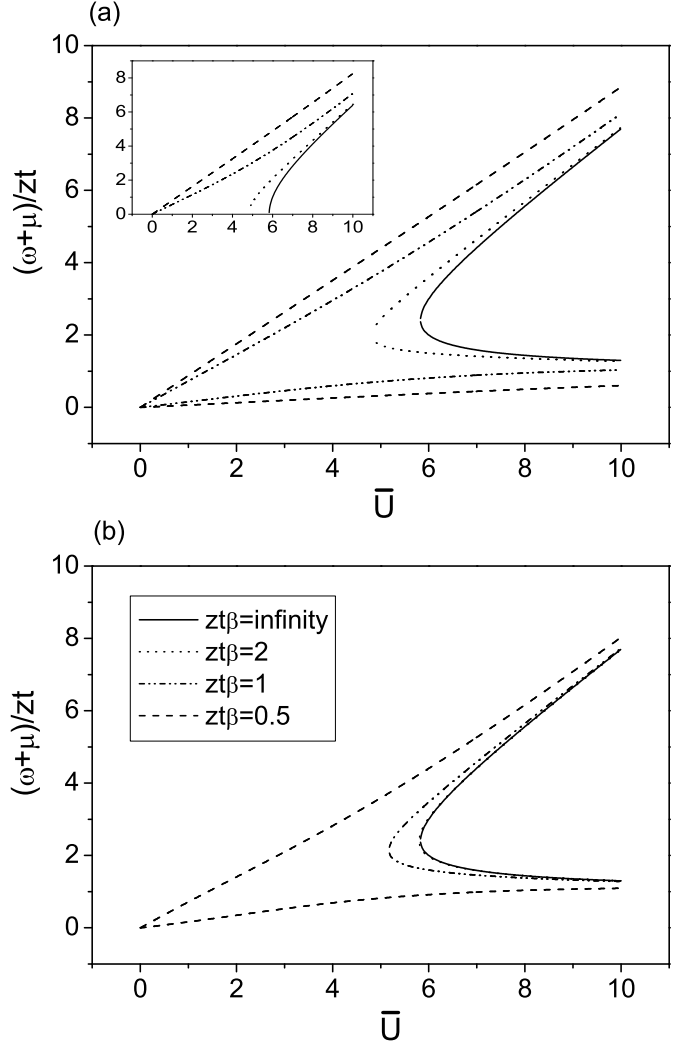}\\
\caption{\label{exci} The excitation energies $(\omega+\mu)/zt$ as
a function of the dimensionless on-site interactions $\bar{U}$ at
$\textbf{k}=0$ for various temperatures. Panel (a) and (b) are
obtained by the slave fermion and the slave boson approaches,
respectively. The inset in panel (a) shows the energy gap between
quasiparticle and quasihole excitations. The average density is
set to be $n=N/L=1$ and three types of slave particles ($n^0$,
$n^1$, and $n^2$) are taken into account.}
\end{center}
\end{figure}

In Fig.\ref{exci}, we show the excitation energies
$(\omega+\mu)/zt$ as a function of $\bar{U}$ at $\textbf{k}=0$ for
various temperatures. We can see that the tip of the lobe, where
the gap for quasiparticle-quasihole excitation disappears, moves
to the small $\bar{U}$ region as the temperature increases. This
picture is qualitatively consistent with the superfluid-normal
phase diagram obtained in Sect. \ref{3}. Another feature of this
figure is that the energy gap for quasiparticle-quasihole
excitation is enlarged as the temperature or the interaction
$\bar{U}$ is increased. Recently, Konabe et al. have obtained the
same result by a standard basis operator method \cite{Konabe}.
When comparing panel (a) with panel (b), we can see that, at the
same temperature, the lobe by the slave fermion evolves away from
zero temperature lobe more greatly than that by the slave boson,
from which we can conclude again that the influence of temperature
on the slave fermion is larger than that on the slave boson. Note
that, except the point of $\bar{U}=0$, the quasiparticle and
quasihole branches do not meet at finite temperature, e.g., the
dotted line in Panel (a) and dashed-dotted-dotted line in Panel
(b) (the gap is very small). This may be due to the finite cut-off
approximation and can be remedied by including more types of slave
particles. In addition, the two branches of spectrum always meet
at (0,0) at high temperature. The reason is that, when the
temperature is high enough, one would have $n^0=n^1=n^2=0.333$ at
$\bar{U}=0$ and then the right hand side of eq.
(\ref{dispersion1}) would always equal zero. In all, eq.
(\ref{dispersion1}) is valid only at very low temperature and near
the Mott-insulator, where the approximation of $\alpha_M=2$ is
justified.

\begin{figure}
\begin{center}
\includegraphics[width=0.6\columnwidth]{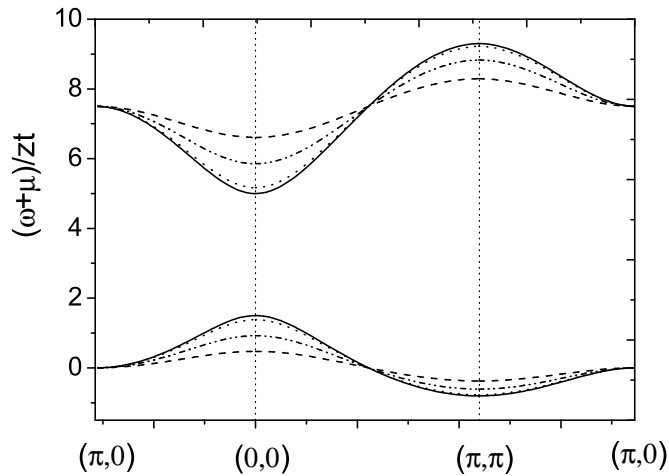}\\
\caption{\label{dispersion} The dispersion relation
$(\omega+\mu)/zt$ of a two-dimensional atomic gas in a square
optical lattice at various temperatures. The results are obtained
by the slave fermion approach and with $\bar{U}=7.5$. The solid,
dotted, dashed-dotted-dotted, and dashed lines correspond to
$zt\beta=$ infinity, 2, 1, and 0.5, respectively.}
\end{center}
\end{figure}

By using the slave fermion approach, we show, in Fig.
\ref{dispersion}, the dispersion relation of a two-dimensional
atomic gas in a square optical lattice at various temperatures.
The behavior of the dispersion curves at zero temperature, marked
by the solid lines, is qualitatively similar to that obtained by
Sengupta et al. \cite{seng} using a strong-coupling expansion
approach. At finite temperature, we can see that, in the different
regions of the Brillouin zone, the influence of the temperature on
the gap for quaiparticle-quasihole excitation is different. In the
region between $(\pi, 0)$ and $(\pi/2, \pi/2)$, the gap increases
as the temperature is increased. However, in the region from
$(\pi/2, \pi/2)$ to $(\pi, 0)$, the gap diminishes when we
increase the temperature. One can see that the temperature affects
the excitation spectrum greatly at the points $(0,0)$ and
$(\pi,\pi)$, but at other points such as $(\pi,0)$ and
$(\pi/2,\pi/2)$ the influence of temperature is very small.

\section{Conclusions}\label{7}

The finite temperature properties of the Bose-Hubbard model were
investigated by both slave boson and slave fermion approaches.
Many physical quantities were calculated in the mean field level
and we found in general there are no qualitatively differences
either by using the slave boson or slave fermion approaches.
However, when studying the stability of the mean field state, we
found that in contrast to slave fermion approach, the slave boson
mean field state is not stable. The mean field phase diagram of
superfluid-normal transition was obtained by making a relative
large $\alpha$ cut-off (e.g. $\alpha_M=6$), and the effect of the
finite types of slave particles was estimated. The unreasonable
local maximum in the phase diagram can be eliminated in the slave
fermion approach by increasing the cut-off, but it can not be
eliminated in the slave boson approach. In the low density region,
the critical temperature at $\bar{U}=0$ by our mean field
approaches is quite close to the Bose-Einstein condensation
temperature of ideal Bose gas in three dimensions. The particle
density was derived and depicted as a function of chemical
potential $\bar{\mu}$. The "Mott-insulating" phase at finite
temperature, where the filling factor is very close to an integer,
was identified and how it evolves as the temperature increases was
demonstrated. The superfluid density was calculated and plotted as
a function of chemical potential $\bar{\mu}$. We showed that the
slave boson approach could not give a correct superfluid density
because the coefficient $a_4$ in the Landau free energy expansion
is always negative. The low-lying excitation spectra were obtained
analytically by taking three types of slave fermions ($n^0$, $n^1$
and $n^2$) into account, and were plotted as a function of
$\bar{U}$ and in the \textbf{k} space at different temperatures.
It was shown that, in the different region of the Brillouin zone,
the influence of the temperature on the quasiparticle-quasihole
gap is different.

\begin{acknowledgments}
One of the authors (X.L.) would like to thank Xiaoyong Feng in the
ITP,CAS for valuable discussions. This work was supported in part
by Chinese National Natural Science Foundation.
\end{acknowledgments}

\begin{appendix}
\setcounter{section}{0} \setcounter{equation}{0}
\renewcommand{\thesection}{\Alph{section}}
\section{the coefficient of fourth order term}\label{appendix a}
In this section we turn to calculate the coefficient $a_4$ of
fourth order term by both slave fermion and slave boson
approaches, following the steps in Ref. \cite{D}. We are only
interested in the $\Phi_{00}$ terms and can write the partition
function of eq. (\ref{action}) as
\begin{eqnarray*}
Z={\int}D\Phi_{00}^*D\Phi_{00}D\bar{a}_\alpha Da_\alpha{e^{-S}},
\end{eqnarray*}
\begin{eqnarray}
S=iL\beta\lambda+\epsilon_0|\Phi_{00}|^2
+\sum_{\alpha\beta}\sum_{\textbf{k},n}
\bar{a}_{\alpha,{\textbf{k}n}}M^{\alpha\beta}a_{\beta,{\textbf{k}n}},
\end{eqnarray}
where $\bar{a}_{\alpha,{\textbf{k}n}}$ and
$a_{\alpha,{\textbf{k}n}}$ are the Grassmann variables in the
slave fermion picture and the ordinary complex numbers in slave
boson picture, $M$ is a matrix:
\begin{gather*}
\begin{pmatrix}\chi_0&A_{0}&&&&\\
B_{0}&\chi_{1}&A_{1}&&&\\
&B_{1}&\chi_2&A_{2}&&\\
&&B_{2}&\chi_3&A_{3}\\
&&&&...&
\end{pmatrix}\quad
\end{gather*}
with
\begin{eqnarray*}
A_{\alpha}&=&-\frac{\sqrt{\alpha+1}}{\sqrt{L\beta}}\epsilon_0\Phi_{00}^*,\\
B_{\alpha}&=&-\frac{\sqrt{\alpha+1}}{\sqrt{L\beta}}\epsilon_0\Phi_{00},\\
\chi_\alpha&=&-i\omega_n+c(\alpha),
\end{eqnarray*}
in which $c(\alpha)$ is defined in eq. (\ref{c(alpha)}). We then
integrate the slave particle fields out of the partition function
and obtain the effective action $S_{eff}'$,
\begin{eqnarray}
Z&=&\int D\Phi_{00}^*D\Phi_{00}e^{-S_{eff}'}\nonumber\\
&=&\int D\Phi_{00}^*D\Phi_{00}
\exp\left\{-\left(iL\beta\lambda+\epsilon_0|\Phi_{00}|^2
\mp\sum_{\textbf{k},n}\ln{[\det\beta{M}]}\right)\right\}\nonumber\\
S_{eff}'&=&iL\beta\lambda+\epsilon_0|\Phi_{00}|^2
\mp\sum_{\textbf{k},n}\ln{[\det\beta{M}]},
\end{eqnarray}
where $\mp$ correspond to slave fermion or slave boson. The
determinant of $\beta{M}$ can be written as
\begin{eqnarray*}
\det\beta{M}&=&\left(\prod_\alpha\beta\chi_\alpha\right)
\left[1-\sum_\alpha\frac{\epsilon_0^2}{L\beta}
\frac{(\alpha+1)}{\chi_\alpha\chi_{\alpha+1}}|\Phi_{00}|^2\right.\\
&&\left.+\sum_\alpha\sum_{|\alpha-\beta|\geq2}\frac{\epsilon_0^4}{(L\beta)^2}
\frac{(\alpha+1)(\beta+1)}{\chi_\alpha\chi_{\alpha+1}\chi_\beta\chi_{\beta+1}}
|\Phi_{00}|^4+\cdots\right].
\end{eqnarray*}
In the case of small $\Phi_{00}$, $\ln{[\det\beta{M}]}$ can be
expanded by using
\begin{eqnarray*}
\ln(1-A|\Phi_{00}|^2+B|\Phi_{00}|^4)=-A|\Phi_{00}|^2+\frac{1}{4}(-2A^2+4B)|\Phi_{00}|^4
    +O(|\Phi_{00}|^5).
\end{eqnarray*}
Then the effective action up to fourth order in $\Phi_{00}$ can be
given by
\begin{eqnarray}\label{effective action}
S_{eff}'&=&iL\beta\lambda\mp\sum_{n}
L\ln\left(\prod_{\alpha}\beta\chi_\alpha\right)
+\left[\epsilon_0\pm\frac{\epsilon_0^2}{\beta}\sum_{n}\sum_\alpha
\frac{(\alpha+1)}{\chi_\alpha\chi_{\alpha+1}}\right]|\Phi_{00}|^2
\\
&\mp&\frac{L}{4}\left(\frac{\epsilon_0}{\sqrt{L\beta}}\right)^4
\sum_{n}\left[-2\left(\sum_\alpha
\frac{(\alpha+1)}{\chi_\alpha\chi_{\alpha+1}}\right)^2
+\sum_\alpha\sum_{|\alpha-\beta|\geq2}
\frac{4(\alpha+1)(\beta+1)}{\chi_\alpha\chi_{\alpha+1}\chi_\beta\chi_{\beta+1}}\right]
|\Phi_{00}|^4,\nonumber
\end{eqnarray}
where the summation over $\textbf{k}$ leads to the number $L$ of
lattice sites. The coefficient $a_2$ of the second-order term
$|\Phi_{00}|^2$ can be read out directly from eq. (\ref{effective
action}),
\begin{eqnarray}
a_2&=&\epsilon_0\pm\frac{\epsilon_0^2}{\beta}\sum_{n}\sum_\alpha
\frac{(\alpha+1)}{\chi_\alpha\chi_{\alpha+1}}\nonumber\\
&=&\epsilon_0-{\epsilon_0}^2\sum_\alpha(\alpha+1)
\frac{n^{\alpha}-n^{\alpha+1}}{-\mu+{\alpha}U} =-G^{-1}(0,0),
\end{eqnarray}
where $\pm$ corresponds to slave fermion or slave boson. We can
see that both slave fermion and slave boson approaches reduce to
the same result $-G^{-1}(0,0)$. The coefficient $a_4$ of the
fourth-order term $|\Phi_{00}|^4$ can be rewritten as
\begin{eqnarray}\label{a4 in appendix}
a_4=\pm\frac{L}{2}\left(\frac{\epsilon_0}{\sqrt{L\beta}}\right)^4
\sum_{n}\sum_\alpha\left[
\frac{(\alpha+1)^2}{\chi_\alpha\chi_\alpha\chi_{\alpha+1}\chi_{\alpha+1}}
+\frac{2(\alpha+1)(\alpha+2)}{\chi_\alpha\chi_{\alpha+1}\chi_{\alpha+1}\chi_{\alpha+2}}\right],
\end{eqnarray}
where $\pm$ corresponds to slave fermion or slave boson.
Typically, we calculate the first term
\begin{eqnarray}\label{the term to calculate}
\pm\frac{1}{\beta}\sum_{n}\left[\frac{(\alpha+1)^2}
{\chi_\alpha\chi_\alpha\chi_{\alpha+1}\chi_{\alpha+1}}\right]
=\pm\frac{1}{\beta}\sum_{n}\frac{(\alpha+1)}{\left[i\omega_n-c(\alpha)\right]^2}
\frac{(\alpha+1)}{\left[i\omega_n-c(\alpha+1)\right]^2}.
\end{eqnarray}
The frequency summations in this term can be performed by using
\begin{eqnarray}
\pm\frac{1}{\beta}\sum_{n}f(i\omega_n)={\rm Res}\left[f(z)\frac{1}
{e^{{\beta}z}\pm1}\right]_{c(\alpha)}
  +{\rm Res}\left[f(z)\frac{1}{e^{{\beta}z}\pm1}\right]_{c(\alpha+1)}.
\end{eqnarray}
Res$\left[F(z)\right]_{c(\alpha)}$ denotes the residue of $F(z)$
at the pole $c(\alpha)$, and for the $m$-th order pole, can be
determined by
\begin{eqnarray}
{\rm
Res}\left[F(z)\right]_{c(\alpha)}=\lim_{z\rightarrow{c(\alpha)}}\frac{1}{(m-1)!}
{\frac{d^{m-1}}{dz^{m-1}}\left[(z-c(\alpha))^mF(z)\right]}.
\end{eqnarray}
By using this equation, the value of eq. (\ref{the term to
calculate}) can be evaluated, which is
\begin{eqnarray}
\pm\frac{1}{\beta}\sum_{n}\left[\frac{(\alpha+1)^2}
{\chi_\alpha\chi_\alpha\chi_{\alpha+1}\chi_{\alpha+1}}\right]
=&-&\frac{(\alpha+1)^2\beta}{(\mu-{\alpha}U)^2}
[n^{\alpha}(1{\mp}n^{\alpha})+n^{\alpha+1}(1{\mp}n^{\alpha+1})]\nonumber\\
&-&\frac{2(\alpha+1)^2}{({\alpha}U-\mu)^3}(n^{\alpha+1}-n^{\alpha}).
\end{eqnarray}
After calculating  the second term in eq. (\ref{a4 in appendix})
in the same way, we can obtain the final expression of $a_4$ as
shown in eq. (\ref{a_4}). In the Mott-insulating region at $T=0$,
we have $n^{\alpha}=\delta_{\alpha,g}$ with $g$ being the average
density. For the slave fermion, the sign in eq. (\ref{a_4}) is
minus and then $a_4$ is reduced to
\begin{eqnarray}
a_4=&-&\frac{{\epsilon_0}^4}{L\beta}\left\{
\frac{g^2}{[(g-1)U-\mu]^3}
+\frac{(g+1)^2}{(\mu-gU)^3}\right.\nonumber\\
&-&\left.\frac{(g+1)(g+2)}{(\mu-gU)^2[2\mu-(2g+1)U]}
-\frac{g(g-1)}{[\mu-(g-1)U]^2[(2g-3)U-2\mu]}\right.\nonumber\\
&-&\left.\frac{g(g+1)U}{[(g-1)U-\mu]^2[\mu-gU]^2} \right\},
\end{eqnarray}
which is exactly the same as the results in Ref. \cite{oosten} and
Ref. \cite{Konabe}. However, for the slave boson, the sign in eq.
(\ref{a_4}) is plus and the result is quite different \cite{note}.
Furthermore, as we have shown in Sec. \ref{5}, $a_4$ is always
non-positive, which leads to the instability of the slave boson
mean field state.
\end{appendix}


\begin{references}
\bibitem{note} We observed several errors in eq. (A9) of Ref. \cite{D}
           in computing the coefficient $a_4$. The most obvious and important one is that they
            took a wrong 'minus' in the first, third, fifth, ninth, and twelfth terms.
\bibitem{note1} In the case of $n=1$,
$\bar{T}_c^{ideal}=2.61^{-2/3}\times4\pi/6\approx1.104$. Note that
the value provided in endnote [15] of Ref. \cite{yy} is slightly
wrong.
\bibitem{note2} One can find a
calculation with $\alpha_M=3$ in Ref. \cite{yy}. The calculation
with $\alpha_M>3$ and whether there exists a reentrant behavior
would be interesting topics for a further work.
\bibitem{note3} The true superfluid density of slave boson can be
obtained by including the sixth order term, but this would predict
a first order superfluid/normal phase transition at some positive
value of $a_2$.
\end{references}
\end{document}